\documentclass[structabstract]{aa}  
\usepackage{graphicx}
\usepackage{txfonts}
\begin{document}
\input{psfig}



\newbox\grsign \setbox\grsign=\hbox{$>$} \newdimen\grdimen \grdimen=\ht\grsign
\newbox\simlessbox \newbox\simgreatbox
\setbox\simgreatbox=\hbox{\raise.5ex\hbox{$>$}\llap
     {\lower.5ex\hbox{$\sim$}}}\ht1=\grdimen\dp1=0pt
\setbox\simlessbox=\hbox{\raise.5ex\hbox{$<$}\llap
     {\lower.5ex\hbox{$\sim$}}}\ht2=\grdimen\dp2=0pt
\def\simgreat{\mathrel{\copy\simgreatbox}}
\def\simless{\mathrel{\copy\simlessbox}}
\newbox\simppropto
\setbox\simppropto=\hbox{\raise.5ex\hbox{$\sim$}\llap
     {\lower.5ex\hbox{$\propto$}}}\ht2=\grdimen\dp2=0pt
\def\simpropto{\mathrel{\copy\simppropto}}

\title{Manganese abundances in Galactic bulge red giants
\thanks{Observations collected at the European  Southern  Observatory,
  Paranal,  Chile  (ESO programmes  71.B-0617A, 73.B0074A, and GTO 71.B-0196)} }
\author{
B. Barbuy\inst{1}
\and
V. Hill\inst {2}
\and
M. Zoccali\inst{3}
\and
D. Minniti\inst{3,4}
\and
A. Renzini\inst{5}
\and
S. Ortolani\inst{6}
\and
A. G\'omez\inst{7}
\and
M. Trevisan\inst{8}
\and
N. Dutra\inst{1}
}
\offprints{B. Barbuy}
\institute{
Universidade de S\~ao Paulo, IAG, Rua do Mat\~ao 1226,
Cidade Universit\'aria, S\~ao Paulo 05508-900, Brazil\\
 e-mail: barbuy@astro.iag.usp.br, nathalia.dutra@usp.br
\and
Universit\'e de Sophia-Antipolis,
 Observatoire de la C\^ote d'Azur, CNRS, Laboratoire Lagrange, BP4229, 06304 Nice Cedex 4, France\\
email: vanessa.hill@oca.eu
\and
Instituto de Astrofisica, Facultad de Fisica, Pontificia Universidad Catolica de Chile, Casilla 306, Santiago 22, Chile\\
e-mail: mzoccali@astro.puc.cl, dante@astro.puc.cl
\and
Vatican Observatory, V00120 Vatican City State, Italy
\and
Osservatorio Astronomico di Padova, Vicolo
 dell'Osservatorio 5, I-35122 Padova, Italy\\
 e-mail:  alvio.renzini@oapd.inaf.it
\and
Universit\`a di Padova, Dipartimento di Astronomia, Vicolo
 dell'Osservatorio 2, I-35122 Padova, Italy\\
 e-mail: sergio.ortolani@unipd.it
\and
Observatoire de Paris-Meudon,  92195 Meudon Cedex, France\\
e-mail: anita.gomez@obspm.fr
\and
Instituto Nacional de Pesquisas Espaciais, Av. dos Astronautas 1758, S\~ao Jos\'e dos Campos 12227-010, Brazil\\
e-mail: marinatrevisan@gmail.com
}

   \date{}

 
  \abstract
   {Manganese is mainly produced in type II SNe during explosive silicon burning,
 in incomplete Si-burning regions, and depends on several nucleosynthesis
environment conditions, such as mass cut beween the matter ejected and falling 
back onto the remnant, electron and neutron excesses, mixing fallback, and
explosion energy. Manganese is also produced in type Ia SNe.}
   {The aim of this work is the study of abundances of
the iron-peak element Mn 
in 56 bulge giants, among which 13 are red clump stars.
Four bulge fields along the minor axis are inspected. The study of 
abundances of Mn-over-Fe as a function of metallicity in the Galactic bulge
may shed light on its production mechanisms.}
   {High-resolution spectra were obtained using the  FLAMES+UVES spectrograph on the Very 
Large Telescope. The spectra were obtained within a program 
to observe 800 stars using the GIRAFFE spectrograph, together with the present UVES spectra.
   }
   {We aim at identifying the chemical evolution of manganese, as a function of metallicity, in the
Galactic bulge. We find  [Mn/Fe] $\sim$-0.7 at [Fe/H]$\sim$-1.3, 
increasing to a solar value at metallicities close to solar,
and showing a spread around
 -0.7$\simless$[Fe/H]$\simless$-0.2, in good agreement with other work 
on Mn in bulge stars.
There is also good agreement with chemical evolution models. We find no clear
 difference in the
behaviour of the four bulge fields. Whereas [Mn/Fe] vs. [Fe/H]
could be identified with the behaviour of the thick disc stars, 
[Mn/O] vs. [O/H]
has a behaviour running parallel, at higher metallicities,
compared to thick disc stars, indicating that the
bulge enrichment might have proceeded differently from that of the thick disk.
}
   {}


   \maketitle
%

\section{Introduction} 
The  Fe-peak elements include (Sc), Ti, V, Cr, Mn, Fe, Co, Ni, Cu, Zn, Ga,
and Ge,
in a broad classification. The lighter iron group includes 
 atomic mass elements in the range
22 $\leq$ Z $\leq$ 26, from  titanium to iron (Woosley \& Weaver 1995,
hereafter WW95).
Despite being gathered in one 
category, iron-peak elements are produced in complex nucleosynthesis processes,
therefore the abundances of some of these elements are
not enriched in locksteps with Fe. 
In particular, the elements Sc, Mn, Cu, and Zn show different
trends to that of Fe (e.g. Nissen et al. 2000; Ishigaki et al. 2013). 
We note that Sc has a behaviour intermediate between that of $\alpha$-elements
and iron-peak elements, and its nucleosynthesis is not well established.

The single manganese isotope $^{55}$Mn is mainly produced in incomplete Si-burning regions, 
during explosive silicon burning in type II SNe (WW95) and in type Ia SNe
(Bravo 2012).
At peak temperature, 4$\times$10$^9$ $<$ T$_{\rm peak}$ $<$ 5$\times$10$^9$ K,
 the shocked material subsequently undergoes explosive nucleosynthesis in
Si burning regions, and unstable $^{55}$Co is produced, which decays into $^{55}$Mn. 
In core collapse SNe the amount of Mn ejected depends on the mass cut between the ejecta
and the inner core (Nakamura et al. 1999). It also depends on
the electron excess Y$_{\rm e}$ (and in turn the progenitor metallicity), 
neutron excess, 
 mixing-fallback process, and explosion energy
(Umeda \& Nomoto 2002, hereafter UN02; 
Kobayashi et al. 2006; Thielemann et al. 1996; WW95). 
Umeda \& Nomoto (2002) considered a Y$_{\rm e}$ larger by only
0.00001 in the incomplete Si-burning region of population III
very low metallicity SNe, and as a result noted that their Mn
yields are smaller by a factor of 10 than the yields of 
Nakamura et al. (1999). These dependencies were also investigated by
Umeda \& Nomoto (2005) at very low metallicities. The production of
Mn in incomplete Si-burning regions in SNe Ia is described in, e.g.
Iwamoto et al. (1999) and  Bravo (2012).

In the present work, we derive Mn abundances for a sample of 56
bulge field stars, observed at high spectral resolution with the 
FLAMES-UVES spectrograph at the Very Large Telescope.
The sample consists of red giants in four bulge fields:
Baade's Window (l=1.14$^{\circ}$, b=-4.2$^{\circ}$), a field at $\rm b=-6^{\circ}$ 
(l=0.2$^{\circ}$, b=$-6^{\circ}$),
the Blanco field (l=0$^{\circ}$, b=$-12^{\circ}$), and a field near NGC 6553
(l=5.2$^{\circ}$, b=$-3^{\circ}$),
as described in Lecureur et al. (2007) and Zoccali et al. (2006; 2008).
The red clump stars in Baade's Window were analysed by Hill et al. (2011). 

We compare our results to previous samples in the literature.
McWilliam et al. (2003a,b) and Sobeck et al. (2006) derived Mn abundances
in seven field and three NGC 6528 bulge giants, respectively.
In order to try to understand the overall Mn chemical enrichment,
the present results are also compared with the trends for
 thin disk, thick disk, and halo stars, derived by Gratton (1989),
Prochaska et al. (2000), Reddy et al. (2003, 2006), Sobeck et al. (2006),
 Feltzing et al. (2007), Nissen \& Schuster (2011), and Cayrel et al. (2004).

We also compare the results with chemical evolution models that take into
account the early enrichment by core collapse SNe and the later type Ia SNe
that cause the increase in [Mn/Fe]\footnote{We adopted here the usual
spectroscopic notation that [A/B] = log(N$_{\rm A}$/N$_{\rm B}$)$_{\star}$ $-$
log(N$_{\rm A}$/N$_{\rm B}$)$_{\odot}$ and $\epsilon$(A) = log(N$_{\rm
A}$/N$_{\rm B}$) + 12 for each elements A and B.}
 towards solar values (Cescutti et al. 2008;
Timmes et al. 1995; Tsujimoto \& Shigeyama 1998).

In Sect. 2 the observations are reported. In Sect. 3 the
atomic constants for the lines under study are given. 
In Sect. 4 the basic stellar parameters are listed, and the
abundance derivation of Mn is described. The results
are derived in Sect. 5 and discussed in Sect. 6. 
Conclusions are drawn in Sect. 7.


\section{Observations}

The spectra were obtained with the FLAMES-UVES spectrograph,
 at the 8.2 m Kueyen ESO telescope.
The mean wavelength coverage is 4800-6800 {\rm \AA}, with a gap at
5775-5825 {\rm \AA}.
The red portion of the spectrum (5800-6800 {\rm \AA}) was obtained with the
ESO CCD \# 20, an MIT backside illuminated, of 4096x2048 pixels, and pixel
size  15x15$\mu$m.
The blue portion of the spectrum (4800-5800 {\rm \AA}) uses ESO Marlene EEV
CCD\#44, backside illuminated, of 4102x2048 pixels, 
and pixel size  15x15$\mu$m. 
With the UVES standard setup 580, the resolution is R $\sim$ 45 000 for a 1
arcsec slit width, and R $\sim$ 55 000 for a slit of 0.8 arcsec. 
The pixel scale is 0.0147 {\rm \AA}/pix, and  
typical signal-to-noise ratios obtained are in the range 
 9 to 70 per pixel, or 25 to 190 per resolution element
($\sim$7.5 pixels at 6000 {\rm \AA}).
 
The spectra were reduced using the FLAMES-UVES 
pipeline\footnote{http://www.eso.org/sci/software/pipelines},
including bias and inter-order background subtraction, flatfield correction,
extraction, and wavelength calibration (Modigliani et al. 2004).
 
The selection of bulge stars was carried out by choosing red giants at
about one magnitude above the horizontal branch, as described in Zoccali et al. (2006).
Red clump stars were selected by Hill et al. (2011).
The list of stars is reported in Sect. 4.

\section{Line parameters: hyperfine structure, oscillator strengths, and solar
abundances}


Hyperfine structure (hfs) in some lines, particularly in odd-Z 
elements, are produced 
from interactions between nuclear and electronic wave functions.
The imbalance of neutrons and protons produces a large nuclear
magnetic moment.

The hfs desaturates strong absorption lines, resulting in
broader lines and higher equivalent widths than if considered as
single lines (e.g. Prochaska \& McWilliam 2000). 
Therefore neglecting hfs effects leads
to overestimating abundances.

Manganese abundances were derived from the \ion{Mn}{I}
 triplet lines at 6000 $\rm \AA$. We also measured the MnI 5394.67  $\rm \AA$ 
line for a few stars, and since this line is strong in giants,
it was used particularly for the metal-poor and/or Mn-poor stars.

Central wavelengths were adopted from the National Institute of
Standards \& Technology (NIST, Martin et al. 2002), which appeared to 
be the most  suitable ones for the location of lines in the solar and Arcturus spectra.

Table \ref{lines2} shows the line excitation potential, log gf
values from the literature, and adopted
total log gf-values for the four studied lines. 
Literature gf-values are from Booth et al. (1984), Blackwell-Whitehead \&
Bergemann (2007, hereafter B-HB07), Den Hartog et al. (2011), and from the line lists
by Kur\'ucz (1993)\footnote{http://kurucz.harvard.edu/atoms.html},
NIST\footnote{http://physics.nist.gov/PhysRefData/ASD/lines$_-$form.html}, 
and VALD (Piskunov et al. 1995).
The adopted gf-values were obtained by fitting the 
solar high-resolution observations using the same UVES spectrograph
as the present sample of spectra
(http://www.eso.org/\-observing/\-dfo/\-quality/\-UVES/\-pipeline/solar{$_{-}$}spectrum.html), 
and Arcturus (Hinkle et al. 2000) spectra. Our astrophysical gf values 
are closer to those of B-HB07.

\begin{figure*}
\centering
\psfig{file=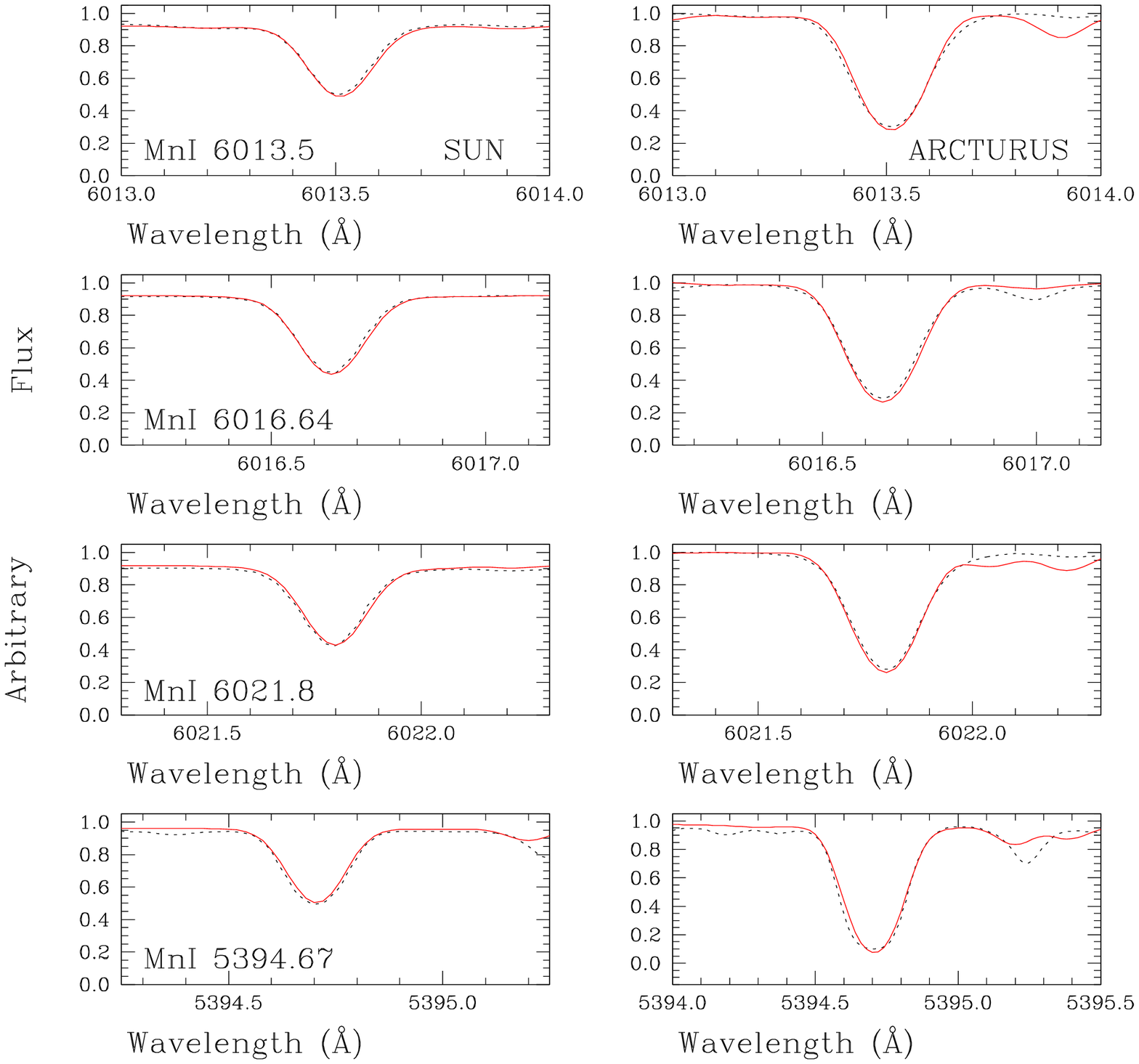,angle=0.,width=14.0 cm}
\caption{Lines of \ion{Mn}{I}  fittings on the solar
spectrum observed with the UVES spectrograph, and on the Arcturus spectrum
 (Hinkle et al. 2000).   
Observed spectrum ({\it dashed lines}); synthetic
spectra ({\it solid red line}).}
\label{sunmn} 
\end{figure*}

The hyperfine structure for the studied lines of \ion{Mn}{I} was taken into account
by employing a code made available by Andrew McWilliam, following the
calculations described by Prochaska \& McWilliam (2000). 
 The nuclear spin (I=2.5) of the only nuclide that
contributes for the manganese abundance ($^{55}$Mn) was found in 
Woodgate \& Martin (1957).
For the \ion{Mn}{I}  6013/6016/6021 and 5394 lines, 
experimental data on hyperfine coupling constants,
namely the magnetic dipole A-factor,
and the electric quadrupole B-factor,
were adopted from Handrich et al. (1969), Biehl (1976), and Brodzinski et al. (1987)
 (transformed from MKaysers(cm-1) to MHz), and are given in Table \ref{lines1}.
The hfs components for the \ion{Mn}{I} lines and
corresponding oscillator strengths are
reported in Table \ref{hfsMn}.
The present fits to the  solar
and Arcturus spectra  for the
\ion{Mn}{I} 
 lines are shown in Fig. \ref{sunmn}.

\begin{table*}
\begin{flushleft}
\caption{Atomic constants used to compute hyperfine structure 
from Handrich et al. (1969), Biehl (1976),
 and Brodzinski et al. (1987). B=0. for 3d$^{5}$4s$^2$ was adopted
when not given in the literature.
}             
\label{lines1}      
\centering          
\begin{tabular}{lrrrrrrrrrrr}     
\noalign{\smallskip}
\hline\hline    
\noalign{\smallskip}
\noalign{\vskip 0.1cm} 
species & $\lambda$ ({\rm \AA}) & Lower level & J & A(MHz) & B(MHz) & Upper level 
& J & A(MHz) & B(MHz)  \\
\noalign{\vskip 0.1cm}
\noalign{\hrule\vskip 0.1cm}
\noalign{\vskip 0.1cm}
MnI & 6013.513 & 3d$^5$4s4p & 1.5  &571.85  & 11.5  & 3d$^5$4s5s & 2.5 & 808.54 & 0. \\
    & 6016.640 & 3d$^5$4s4p & 2.5 & 466.94 & $-$72.93 & 3d$^5$4s5s & 2.5 & 808.54 & 0.  \\ 
 & 6021.800 & 3d$^5$4s4p & 3.5  & 429.165& 66.6 &3d$^5$4s5s & 2.5 & 808.54 & 0.  \\
 & 5394.669 & 3d$^5$4s$^2$ & 2.5 &$-$71.950 &0. &3d$^5$4s4p & 3.5 &545.622 &101.930 \\
\noalign{\vskip 0.1cm}
\noalign{\hrule\vskip 0.1cm}
\noalign{\vskip 0.1cm}  
\hline                  
\end{tabular}
\end{flushleft}
\end{table*}  

\begin{table*}
\begin{flushleft}
\caption{Central wavelengths and total literature and adopted oscillator strengths.}             
\label{lines2}      
\centering          
\begin{tabular}{lrrrrrrrrrrrrrr}     
\noalign{\smallskip}
\hline\hline    
\noalign{\smallskip}
\noalign{\vskip 0.1cm} 
species & {\rm $\lambda$} ({\rm \AA}) & {\rm $\chi_{\rm ex}$ (eV)} 
& {\rm gf$_{\rm Booth}$} & {\rm gf$_{\rm Kurucz}$} &
 {\rm gf$_{\rm NIST}$} & {\rm gf$_{\rm VALD}$}  
& {\rm gf$_{\rm B-HB07}$} & {\rm gf$_{\rm DenHartog}$}   & {\rm gf$_{\rm adopted}$}  \\
\noalign{\vskip 0.1cm}
\noalign{\hrule\vskip 0.1cm}
\noalign{\vskip 0.1cm}
MnI & 6013.513 & 3.072451 & $-$0.251 & $-$0.251 & $-$0.252 &$-$0.252 & -0.43 & $-$0.352 & $-$0.40 \\
    & 6016.640 & 3.073534 & ---& $-$0.216 & --- & 0.0 & $-$0.183 & -0.25 & $-$0.216 \\ 
    & 6021.800 & 3.075294 & 0.034 & 0.034 & 0.035 & 0.035 & -0.054 & -0.12 & -0.10 \\
    & 5394.67 & 0.0     & $-$3.503 & $-$3.591 & $-$3.503 & $-$3.503 & -- & -- & $-$3.55 \\
\noalign{\vskip 0.1cm}
\noalign{\hrule\vskip 0.1cm}
\noalign{\vskip 0.1cm}  
\hline                  
\end{tabular}
\end{flushleft}
\end{table*}


\begin{table*}
\caption{Hyperfine structure for \ion{Mn}{I} lines. }
\label{hfsMn}
\centering
\begin{tabular}{cccccccccccc}
\hline
\noalign{\smallskip}
\multicolumn{2}{c}{6013.488$\rm \AA$;  $\chi$=3.072 eV} && \multicolumn{2}{c}{6016.673$\rm \AA$; $\chi$=3.073 eV}
 && \multicolumn{2}{c}{6021.792$\rm \AA$; $\chi$=3.075 eV} 
&& \multicolumn{2}{c}{5394.669$\rm \AA$;  $\chi$=0.0 eV} \\
\multicolumn{2}{c}{log gf(total) = $-$0.40} && \multicolumn{2}{c}{log
gf(total) = $-$0.266} && \multicolumn{2}{c}{log gf(total) = $-$0.10}
&& \multicolumn{2}{c}{log gf(total) = $-$3.55}  \\
\noalign{\smallskip}
\cline{1-2} \cline{4-5} \cline{7-8} \cline{10-11} \\
$\lambda$ ($\rm \AA$) & log gf && $\lambda$ ($\rm \AA$) & log gf
&& $\lambda$ ($\rm \AA$) & log gf
&& $\lambda$ ($\rm \AA$) & log gf \\
\noalign{\smallskip}
\cline{1-2} \cline{4-5} \cline{7-8} \cline{10-11} \\
6013.559 & $-$1.9563 && 6016.665 & $-$1.8223 && 6021.829 & $-$1.6563 
&& 5394.722 & $-$5.1063 \\
6013.549 & $-$1.6341 && 6016.681 & $-$1.8223 && 6021.819 & $-$1.6263
&& 5394.722 & $-$5.0763 \\
6013.529 & $-$1.8102 && 6016.671 & $-$2.5882 && 6021.799 & $-$2.4045
&& 5394.712 & $-$4.8211  \\
6013.563 & $-$2.0021 && 6016.651 & $-$1.5602 && 6021.828 & $-$1.3711
&& 5394.724 & $-$5.8545 \\
6013.543 & $-$1.4792 && 6016.683 & $-$1.5602 && 6021.809 & $-$1.4502
&& 5394.714 & $-$4.9002  \\
6013.514 & $-$1.4000 && 6016.664 & $-$1.8892 && 6021.779 & $-$2.3253
&& 5394.699 & $-$4.5992  \\
6013.564 & $-$2.1782 && 6016.634 & $-$1.4579 && 6021.823 & $-$1.1492
&& 5394.716 & $-$5.7753  \\
6013.535 & $-$1.4580 && 6016.681 & $-$1.4579 && 6021.794 & $-$1.3833 
&& 5394.701 & $-$4.8333  \\
6013.496 & $-$1.1270 && 6016.652 & $-$1.4421 && 6021.755 & $-$2.4502
&& 5394.681 & $-$4.4152  \\
6013.563 & $-$2.5584 && 6016.613 & $-$1.4633 && 6021.814 & $-$0.9652
&& 5394.704 & $-$5.9002 \\
6013.524 & $-$1.6041 && 6016.675 & $-$1.4633 && 6021.775 & $-$1.4089
&& 5394.684 & $-$4.8589  \\
6013.475 & $-$0.9149 && 6016.636 & $-$1.1111 && 6021.726 & $-$2.8024 
&& 5394.657 & $-$4.2568  \\
         &         && 6016.587 & $-$1.6260 && 6021.801 & $-$0.8068
&& 5394.687 & $-$6.2524  \\
         &         && 6016.663 & $-$1.6260 && 6021.752 & $-$1.5849 
&& 5394.661 & $-$5.0349 \\
         &         && 6016.614 & $-$0.8479 && 6021.785 & $-$0.6673 \\
\noalign{\vskip 0.1cm}
\noalign{\hrule\vskip 0.1cm}
\noalign{\vskip 0.1cm}  
\hline   
\end{tabular}
\end{table*}


The adopted solar abundances for Fe and Mn are from Grevesse \& Sauval
(1998): ($\rm log \epsilon (Fe)_{\odot}$, log $\epsilon (Mn)_{\odot}$) = (7.50, 5.39),
in good agreement with those from Asplund et al. (2009) and Lodders et
al. (2009), of (7.50, 5.43) and (7.46, 5.50) respectively.

For Arcturus we adopted the parameters from Mel\'endez et al. (2003)
of (T$_{\rm eff}$(K), log g, [Fe/H], v$_{\rm t}$(km.s$^{-1}$) = (4275,
1.55, $-$0.54, 1.65), which are in good agreement with those
 from Ram\'{\i}rez \&Allende Prieto (2011) of (4286, 1.66, $-$0.52, 1.74).
For Arcturus, a Mn abundance of [Mn/Fe]=-0.25 fitted the Mn lines
with the same parameters as for the Sun; this is in excellent agreement with
the Mn abundance given by Ram\'{\i}rez \&Allende Prieto (2011) of
[Mn/Fe] = -0.21.

\begin{figure}
\centering
\psfig{file=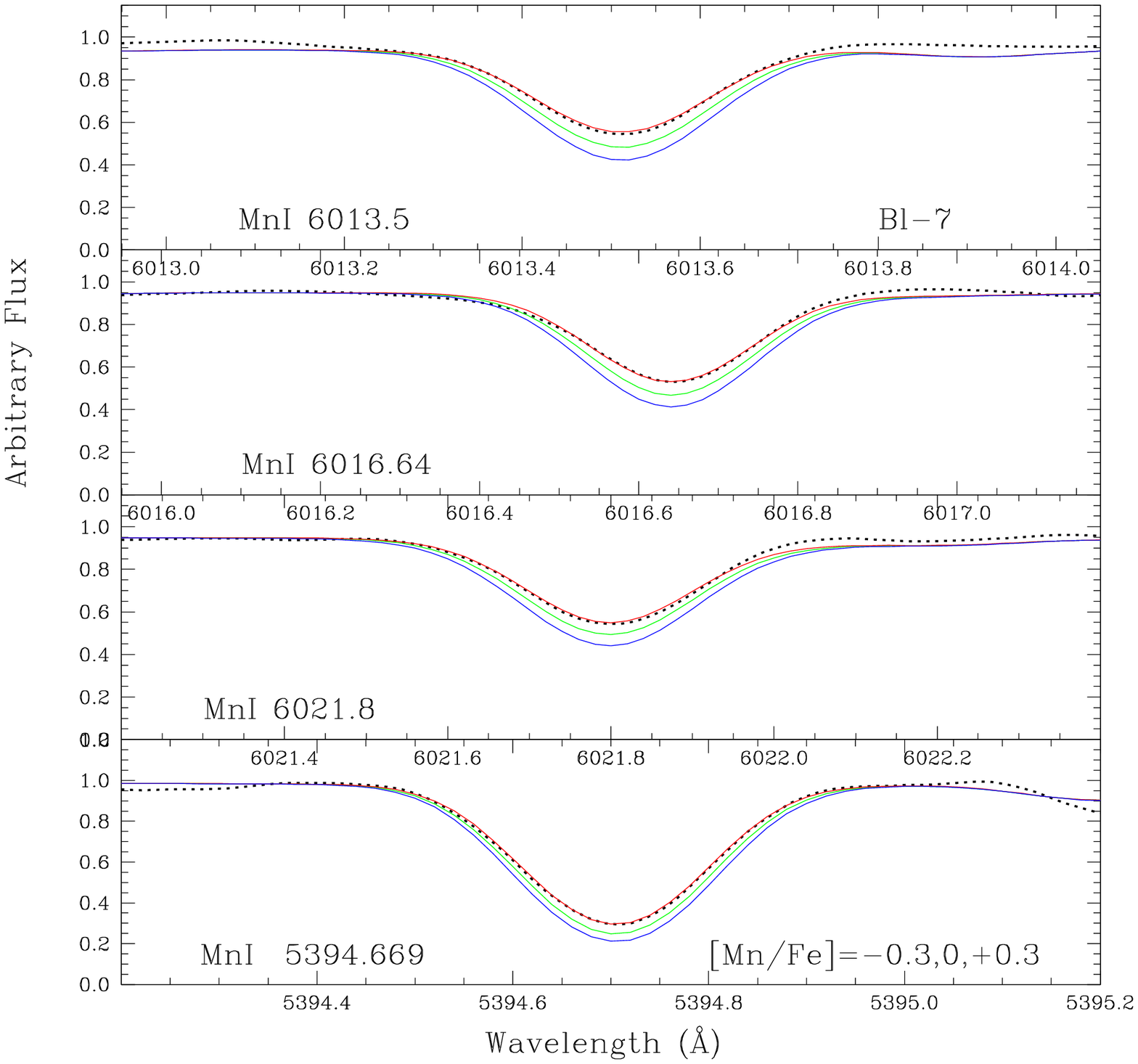,angle=0.,width=9.0 cm}
\caption{Fits of the four \ion{Mn}{I} lines for the metal-poor ([Fe/H]=-0.47) star BL-7.}
\label{bl7} 
\end{figure}

\begin{figure}
\centering
\psfig{file=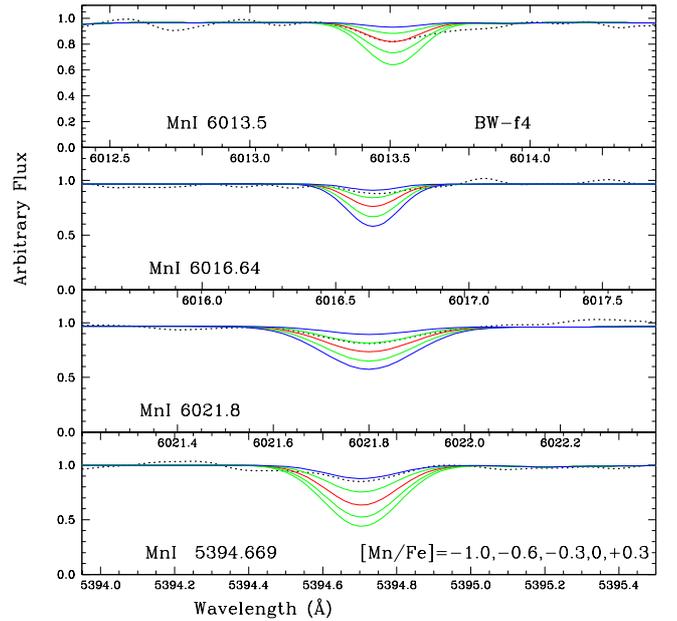,angle=0.,width=9.0 cm}
\caption{Fits of the four \ion{Mn}{I} lines for the metal-poor ([Fe/H]=-1.21) star BW-f4}.
\label{bwf4} 
\end{figure}

\begin{figure}
\centering
\psfig{file=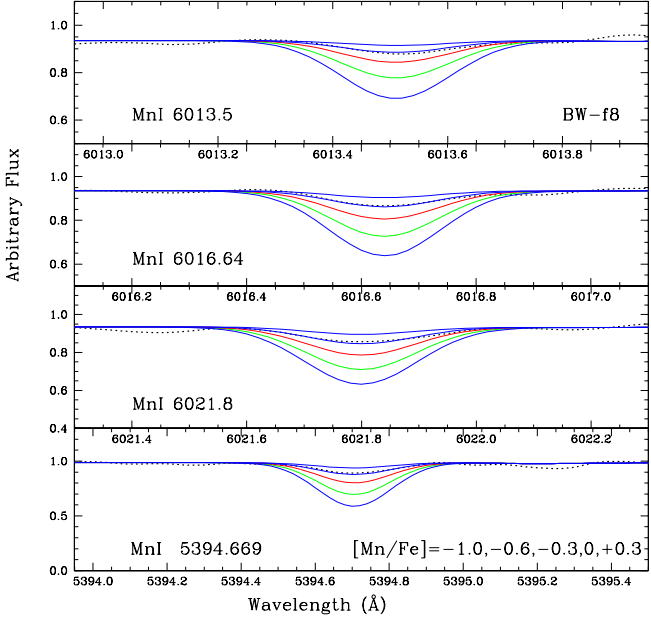,angle=0.,width=9.0 cm}
\caption{Fits of the four \ion{Mn}{I} lines for the most metal-poor ([Fe/H]=-1.27) sample star BW-f8.}
\label{bwf8} 
\end{figure}

\begin{figure}
\centering
\psfig{file=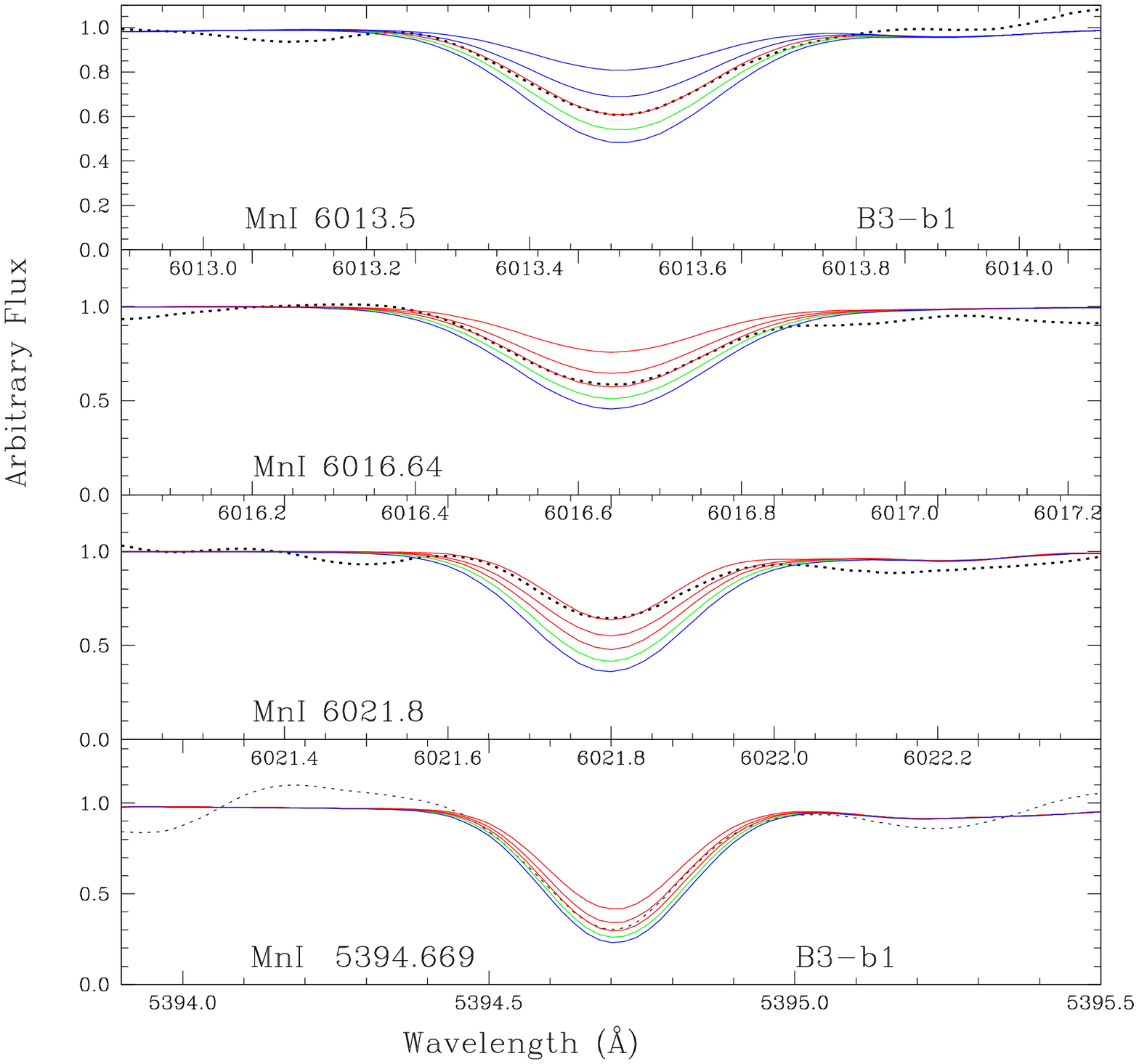,angle=0.,width=9.0 cm}
\caption{Fits of the four \ion{Mn}{I} lines for the metal-poor ([Fe/H]=-0.78) star B3-b1}.
\label{b3b1} 
\end{figure}
\begin{figure}
\centering
\psfig{file=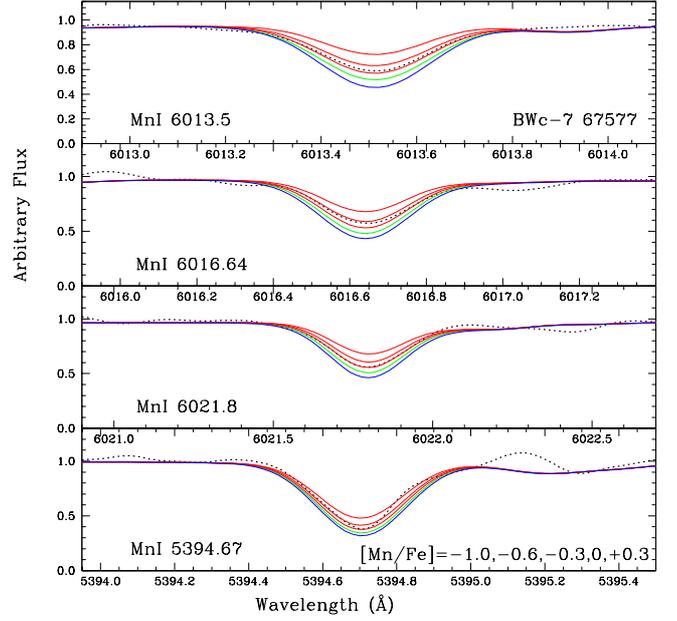,angle=0.,width=9.0 cm}
\caption{Fits of the four \ion{Mn}{I} lines for the ([Fe/H]=-0.25) red clump star BWc-7}.
\label{bwc7} 
\end{figure}

\begin{figure*}
\centering
\psfig{file=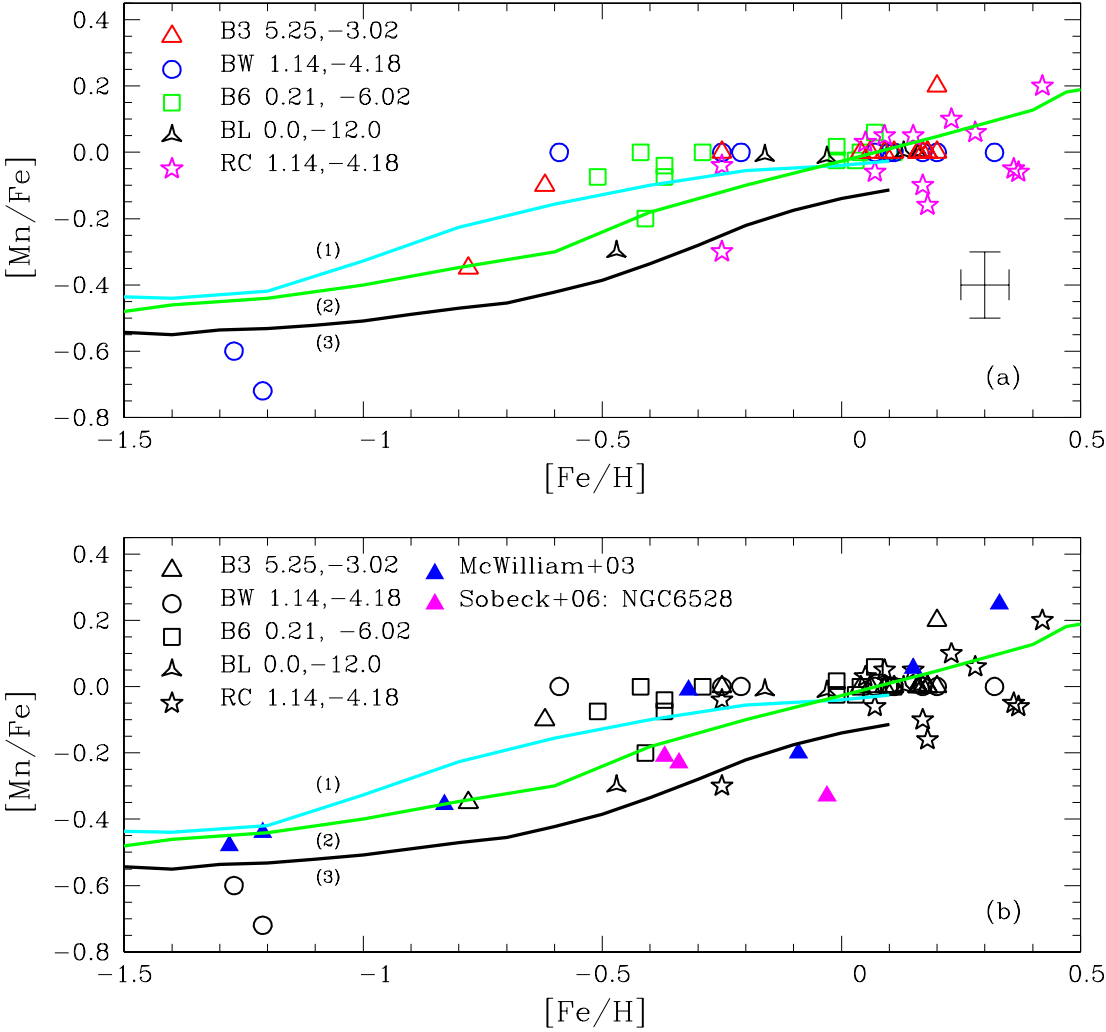,angle=0.,width=15.0 cm}
\caption{[Mn/Fe] vs. [Fe/H]: (a) present results for the four bulge fields.
Symbols: open red triangles: B3 stars from the NGC 6553 field
(l=5.2$^{\circ}$, b=$-3^{\circ}$); open blue circles: BW stars from
Baade's Window (l=1.14$^{\circ}$, b=-4.2$^{\circ}$);
open green squares: B6 stars from a field at $\rm b=-6^{\circ}$ 
(l=0.2$^{\circ}$, b=$-6^{\circ}$); open black triangles:
Bl stars from the Blanco field (l=0$^{\circ}$, b=$-12^{\circ}$);
open magenta stars: red clump (RC) stars from Baade's Window.
An error bar is given in the bottom-right corner of Fig. 2a.
(b) same as (a), here including 
bulge field stars (full blue triangles) 
from McWilliam et al. (2003a,b), and NGC 6528
from Sobeck et al. (2006) (full magenta triangles).
Solid lines correspond to chemical evolution models by (1) TS98 (blue),
(2) Cescutti et al. (2008) (green), and (3) Timmes et al. (1995) (black).
}
\label{plotbojo} 
\end{figure*}

 \begin{figure}
\centering
\psfig{file=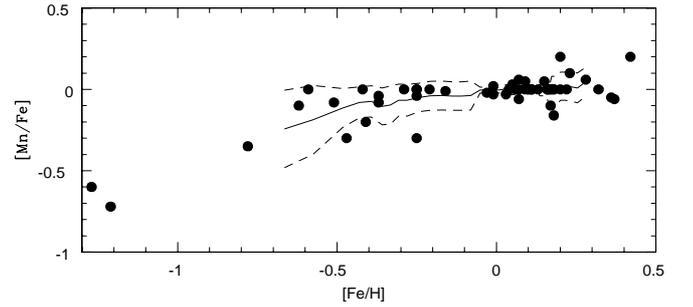,angle=0.,width=9.0 cm}
\caption{Present results on [Mn/Fe] vs. [Fe/H]
together with a moving average line (moving average on 10 points) 
and the corresponding rms dispersion in each subsample 
(shown as dashed lines).
}
\label{movingrms} 
\end{figure}

\begin{figure*}
\centering
\psfig{file=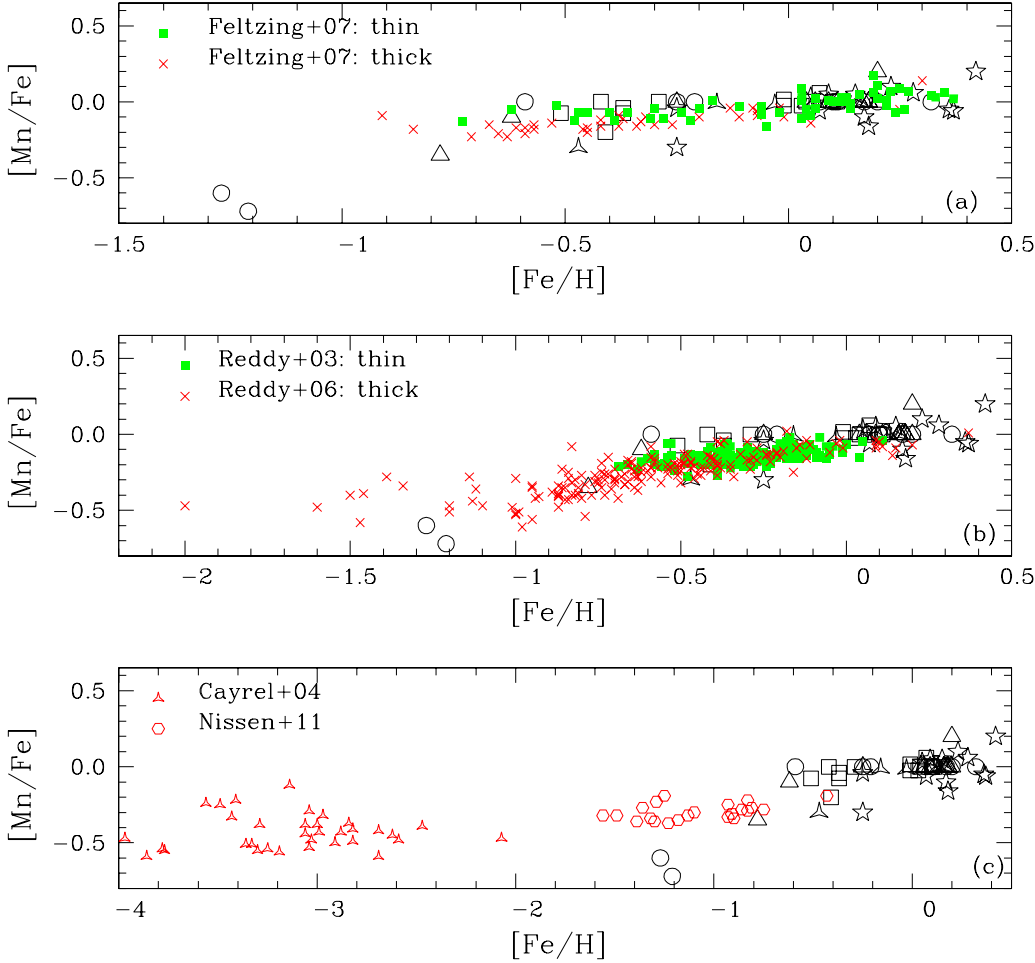,angle=0.,width=14.0 cm}
\caption{[Mn/Fe] vs. [Fe/H]: Present results, shown with the same symbols as in 
Fig. \ref{plotbojo},
compared with the literature: a) Feltzing et al. (2007) and b) Reddy et al. (2003, 2006)
 for thin disc (full green squares) and thick disc (red crosses) stars; 
c) Cayrel et al. (2004) (open red triangles) and Nissen \& Schuster (2011) (open red circles)
for halo stars. }
\label{lit} 
\end{figure*}

\begin{figure*}
\centering
\psfig{file=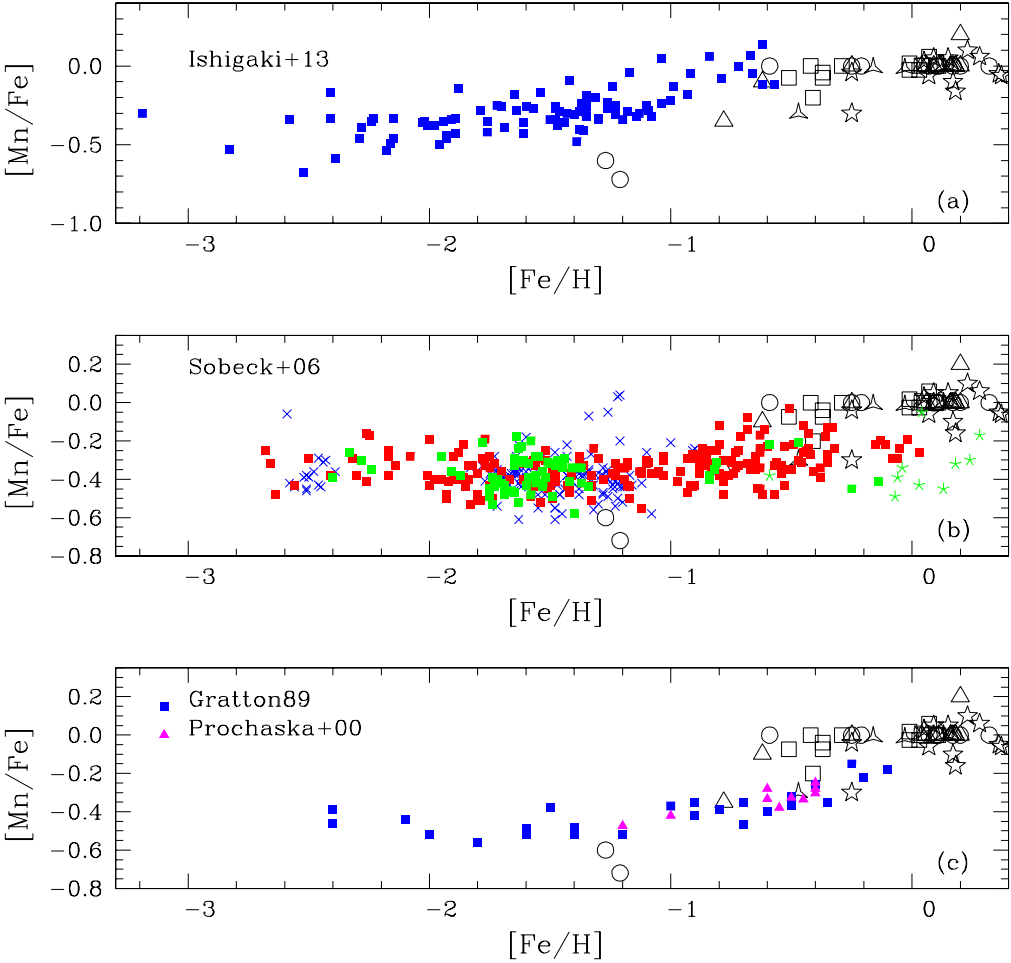,angle=0.,width=14.0 cm}
\caption{[Mn/Fe] vs. [Fe/H]: Present results,   with the same symbols as in Fig. \ref{plotbojo}, compared with
a) Ishigaki et al. (2013): thick disc and halo (blue squares);
b) Sobeck et al. (2006): globular clusters (blue crosses); 
field stars (red squares),  
bulge metal-rich globular clusters plus the old open cluster Cr261 with stellar parameters
from the literature (green stars);
c) Gratton (1989) (blue squares), and Prochaska et al. (2000) (thick disc: full magenta triangles).}
\label{plotsobeck} 
\end{figure*}

\begin{figure*}
\centering
\psfig{file=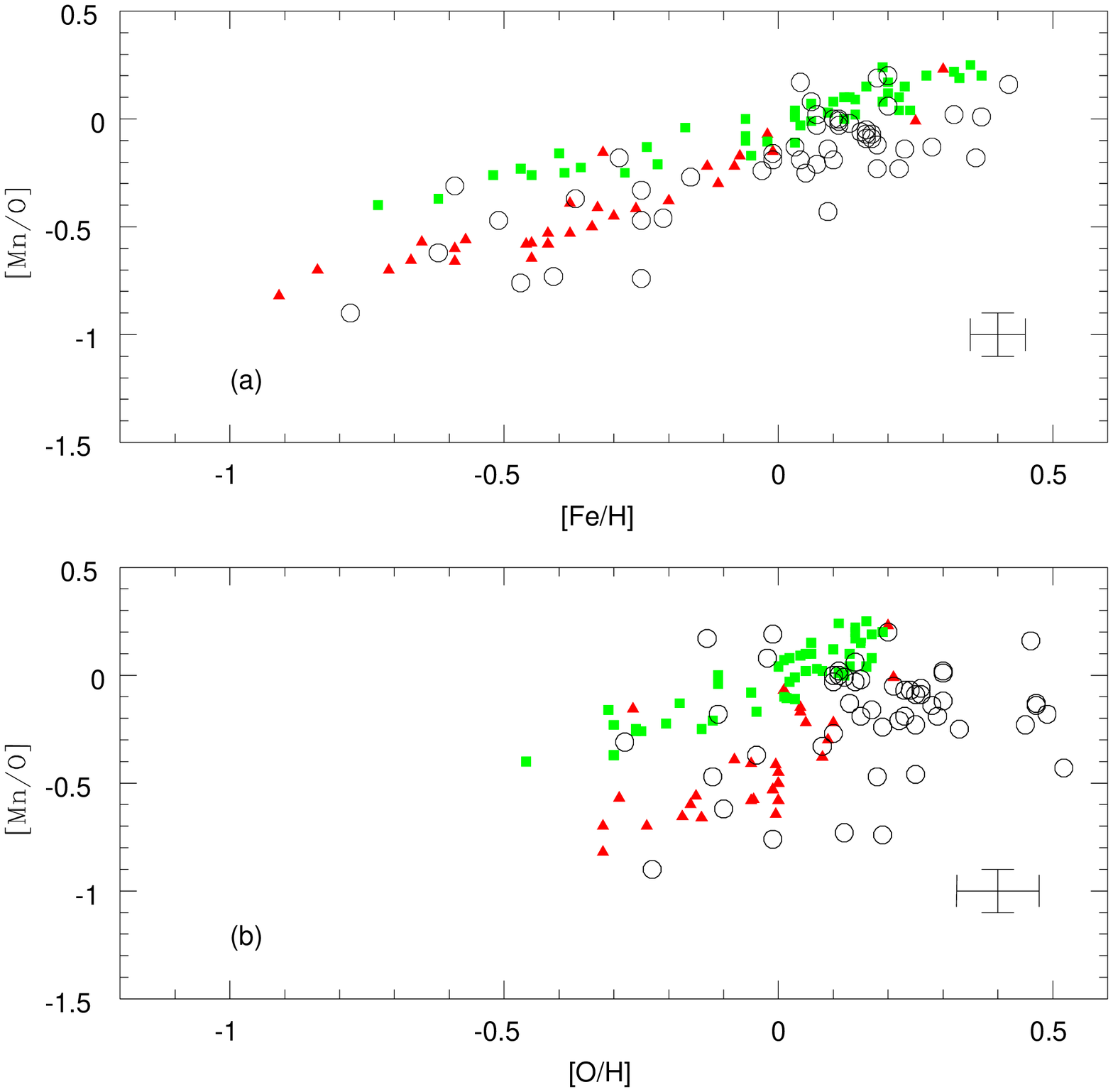,angle=0.,width=14.0cm}
\caption{a) [Mn/O] vs. [Fe/H]; b) [Mn/O] vs. [O/H]: Present results (open circles),
compared with results from Feltzing et al. (2007)
 with same symbols as in Fig. \ref{lit}a for 
thin disc (green squares) and thick disc (red full triangles) stars.
Errorbars are given on the bottom-right corner of both panels.
}
\label{mno1} 
\end{figure*}



\section{Abundance analysis}

The list of stars is reported in Table \ref{atmos1},
 together with their stellar parameters. 

\subsection{Atmospheric parameters and abundance calculation}
Manganese abundances were obtained through line-by-line spectrum synthesis
calculations. The calculations of synthetic spectra were carried out 
using the code PFANT described in Barbuy et al. (2003).
Molecular lines of the  CN  A$^2$$\Pi$-X$^2$$\Sigma$, C$_2$  Swan 
A$^3$$\Pi$-X$^3$$\Pi$, and TiO A$^3$$\Phi$-X$^3$$\Delta$ $\gamma$ and
B$^3$$\Pi$-X$^3$$\Delta$ $\gamma$' systems are taken into account.

The MARCS Local Thermodynamic Equilibrium (LTE) 1D model atmospheres grid was employed
 (Gustafsson et al. 2008).
 For the red giant stars, we used the the parameters derived by Zoccali et al. (2006) 
and Lecureur et al. (2007), while for red clump stars we used the parameters updated by Hill et al. (2011).
For these stars, Hill et al. (2011) carried out the analyses based both on
GIRAFFE spectra and UVES spectra of the same stars, and we adopted the UVES
stellar parameters.


Gaussian convolutions with $\sigma$=2, 3, and 4, and even 5 pixels for the noisier spectra, 
with 0.0147 {\rm \AA}/pixel,
 were applied to all
spectra, to improve the fits to the faint lines studied in this work.

For the few metal-poor stars, we added the fit to
 the MnI 5394 {\rm \AA} line, in order to
be able to derive their Mn abundances more accurately.
The fits to the four MnI lines
for metal-poor sample stars are shown in Figs. \ref{bl7}-\ref{b3b1}.
Mn abundances are reported in Table \ref{atmos1}, where Mn1\-/Mn2\-/Mn3\-/Mn4 refer to
the 6013\-/6016\-/6021\-/5934 {\rm \AA} lines.

\subsection{Uncertainties}
The main uncertainties are those coming from the derivation of stellar parameters,
as described in Lecureur et al. (2007), where the following uncertainties were
adopted: $\pm$ 200 K for
temperature, $\pm$ 0.20 for surface gravity 
 (photometric gravities derived assuming that the stars are at the distance 
of the bulge),
$\pm$0.1 for metallicity,  and $\pm$ 0.10 kms$^{-1}$ for
microturbulence velocity. The profile fitting errors are to be added to these,
estimated to be of the order of $\pm$0.1 in Mn-over-Fe abundances.
For eight sample stars, Ryde et al. (2010) carried out a re-analysis using the same
equivalent widths measured by Lecureur et al. (2007). They found cooler temperatures
by about 160 K, and estimated errors in spectroscopic gravity of $\pm$0.5. 
Hill et al. (2011) reported errors in [Fe/H] star-by-star for the GIRAFFE spectra.
Based on the uncertainties on stellar parameters as discussed in Lecureur et al. (2007)
and Ryde et al. (2010), we assumed uncertainties
 of $\pm$ 150 K in temperature, $\pm$0.2 in log g 
given that we used photometric gravities, 
$\pm$0.1 in [Fe/H], and $\pm$0.1 in microturbulence velocity.
Table \ref{uncertainties} reports the errors in [Mn/Fe] due to a change in the parameters
for two giants close to the edges of our [Fe/H] range:
 the metal-poor star B3-B1 and the red clump star BWc-6.
In this table we also report the different results
by adopting the stellar parameters from the several analyses carried out
as described above.
The errors in [Mn/Fe] are computed using model atmospheres corresponding
to each of these models, and rederiving [Mn/Fe] from
 fitting the observed spectra. The errors are the differences relative to
the fitting with the adopted model. 
 The overall uncertainties due to stellar parameters
 are given in Table \ref{errors2} for these two sample stars.
We note that these final errors
are upper limits, given that the stellar parameters are covariant.
Finally,  adding $\pm$0.1 
due to a continuum location uncertainty, we estimate a mean uncertainty in 
[Mn/Fe] values of $\pm$0.2.

Non-LTE effects on Mn lines were computed by Bergemann \& Gehren (2008). While the effects
are considerable in metal-poor stars hotter than 5000 K, increasing with increasing temperature
and decreasing metallicity, they appear to be below (non-LTE-LTE)$\simless$+0.1dex for the 
metallicities and temperatures of the present sample stars. We preferred not to
try to correct the literature Mn abundances for halo stars, given that non-LTE effects
are counterbalanced by 3D modelling, therefore a correction taking into account both
effects is not known.

\begin{table*}
\begin{flushleft}
\caption{Atmospheric parameters adopted from Zoccali et al. (2006) and Lecureur
et al. (2007) for the red giants, and from Hill et al. (2011) for the red clump stars, 
and abundances of [Mn/Fe] derived in the present work.
Mn1/Mn2/Mn3/Mn4 correspond to the 6013/6016/6021/5394 {\rm \AA}
lines. }             
\label{atmos1}      
\centering          
\begin{tabular}{l@{}rr@{}rr@{}r@{}r@{}r@{}c@{}r@{}r@{}r@{}r@{}r@{}r@{}}     
\noalign{\smallskip}
\hline\hline    
\noalign{\smallskip}
\noalign{\vskip 0.1cm} 
Star & OGLE~n$^{\circ}$ & $\phantom{-}\alpha$(J2000) & $\delta$(J2000) &  $V$ &
  $\phantom{-}$$\phantom{-}$T$_{\rm eff}$ & $\phantom{-}$log~$g$ & $\phantom{-}$[Fe/H] &  
 $\phantom{-}$v$_{\rm t}$ &
 [Mn1/Fe] & [Mn2/Fe] & [Mn3/Fe] & [Mn4/Fe] & [Mn/Fe] & \\                            
\noalign{\vskip 0.1cm}
\noalign{\hrule\vskip 0.1cm}
 &  & (h,m,s)  & (${\circ}$,~\arcmin,~\arcsec)& [mag] & \phantom{-}[K] &  & & [kms$^{-1}$] & & & & & \hbox{mean}& \\
\noalign{\vskip 0.1cm}    
\noalign{\vskip 0.1cm}
\noalign{\hrule\vskip 0.1cm}
\noalign{\vskip 0.1cm}  
B6-b1 & 29280c3 & 18 09 50.604 & \phantom{-}$-$31 40 51.599 &16.14& 4400 & 1.8 & 0.07 & 1.6 & 0.20 & 0.00 & 0.00 & -- &  0.06 &  \\
B6-b2 & 83500c6 & 18 10 34.117 & \phantom{-}$-$31 49 09.094&16.40&4200 &1.5 &$-$0.01 &1.4 &--- & $-$0.05 & 0.00 & -- &$-$0.03 &  \\
B6-b3 & 31220c2 & 18 10 19.183 & \phantom{-}$-$31 40 28.102 &16.09& 4700 & 2.0 & 0.10 & 1.6 & 0.00 &  0.00 &  0.00 &--&  0.00 &  \\  
B6-b4 & 60208c7 &  18 10 07.901 & \phantom{-}$-$31 52 41.288&16.12& 4400 & 1.9 & $-$0.41 & 1.7 & 0.00   & $-$0.25 & $-$0.25 &$-$0.30 & $-$0.20 & \\
B6-b5 & 31090c2 & 18 10 37.509 & \phantom{-}$-$31 40 29.098&16.04& 4600 & 1.9 & $-$0.37 & 1.3 & 0.00   & $-$0.05  & 0.00  & $-$0.10 & $-$0.04 &  \\
B6-b6 & 77743c7 & 18 09 49.217 & \phantom{-}$-$31 50 07.597&16.09&4600 & 1.9 & 0.11 & 1.8 & 0.00 & 0.00 & 0.00&-- & 0.00  &   \\
B6-b8 & 108051c7 & 18 09 56.070 & \phantom{-}$-$31 45 46.301&16.29& 4100 & 1.6 & 0.03 & 1.3 & -- & $-$0.05 & 0.00 & -- & $-$0.03 & \\

B6-f1 & 23017c3 & 18 10 04.591 & \phantom{-}$-$31 41 45.295&15.96& 4200 & 1.6 & $-$0.01 & 1.5 & 0.05 & 0.00 & 0.00&-- & 0.02 & \\
B6-f2 & 90337c7 & 18 10 11.636 & \phantom{-}$-$31 48 19.196&15.91& 4700 & 1.7 & $-$0.51 & 1.5 & 0.00   & 0.00  & 0.00 & $-$-0.30 & $-$0.08 &  \\
B6-f3 & 21259c2 & 18 10 17.850 & \phantom{-}$-$31 41 55.196&15.71& 4800 & 1.9 & $-$0.29 & 1.3 &0.00  & 0.00  & 0.00 &0.00 & 0.00 & \\
B6-f5 & 33058c2 & 18 10 41.643 & \phantom{-}$-$31 40 11.801&15.90& 4500 & 1.8 & $-$0.37 & 1.4 & 0.00   & 0.00  & 0.00 &$-$0.30 & $-$0.08 &\\
B6-f7 & 100047c6 & 18 10 52.430 & \phantom{-}$-$31 46 42.097&15.95& 4300 & 1.7 & $-$0.42 & 1.6 &---   & 0.00  & 0.00 &-- & 0.00 & \\
B6-f8 & 11653c3 & 18 09 56.963 & \phantom{-}$-$31 43 22.503&15.65&4900 & 1.8 & 0.04 & 1.6 &  0.00  &0.00  & 0.00  &-- & 0.00 & \\

BW-b2 & 214192 & 18 04 24.077 & \phantom{-}$-$30 05 57.797&16.58& 4300 & 1.9 & 0.22 & 1.5 & 0.00 & 0.00 & 0.00 &-- & 0.00 & \\
BW-b4 & 545277 & 18 04 05.476 & \phantom{-}$-$30 05 52.496&16.95& 4300 & 1.4 & 0.07 & 1.4 & 0.00 & 0.00 & 0.00 &-- &  0.00 & \\
BW-b5 & 82760 & 18 04 13.407 & \phantom{-}$-$29 58 17.800&16.64& 4000 & 1.6 & 0.17 & 1.2 & 0.00  & 0.00 &0.0  &-- & 0.0  &  \\
BW-b6 & 392931 & 18 03 51.969 & \phantom{-}$-$30 06 27.900&16.42& 4200 & 1.7 & $-$0.25 & 1.3 & 0.00   & 0.00  & 0.00 &$-$0.50: & 0.00 & \\
BW-b7 & 554694 & 18 04 04.693 & \phantom{-}$-$30 02 39.604&16.69& 4200 & 1.4 & 0.10 & 1.2 & 0.00 & --  & 0.00 &-- & 0.00 &  \\

BW-f1 & 433669 & 18 03 37.268 & \phantom{-}$-$29 54 22.294&16.14& 4400 & 1.8 & 0.32 & 1.6 &0.00 & 0.00 & 0.00 &-- & 0.00 & \\
BW-f4 & 537070 & 18 04 01.528 & \phantom{-}$-$30 10 20.700&16.07&4800 & 1.9 &  $-$1.21 & 1.7 & $-$0.30  & $-$0.80 & $-$0.60 &$-$0.90 & $-$0.72 &  \\
BW-f5 & 240260 & 18 04 39.753 & \phantom{-}$-$29 55 19.794&15.88& 4800 & 1.9 & $-$0.59 & 1.3 &  0.00   & 0.00  & 0.00 &$-$0.50: & 0.00 & \\
BW-f6 & 392918 & 18 03 37.014 & \phantom{-}$-$30 07 04.299&16.37& 4100 & 1.7 & $-$0.21 & 1.5 &  0.00   & 0.00  & 0.00 &$-$0.50: & 0.00 & \\
BW-f7 & 357480 & 18 04 44.045 & \phantom{-}$-$30 03 15.193&16.31& 4400 & 1.9 & 0.11 & 1.7 & --- & 0.00 & 0.00 & -- & 0.00 & \\
BW-f8 & 244598 & 18 03 30.615 & \phantom{-}$-$30 01 44.803&16.00& 5000 & 2.2 & $-$1.27 & 1.8 & $-$0.60  & $-$0.60 & $-$0.60 &$-$0.60 & $-$0.60& \\

BL-1 & 1458c3 & 18 34 58.643& \phantom{-}$-$34 33 15.241 & 15.37 & 4500 & 2.1 & $-$0.16 & 1.5 &  0.00 & $-$0.03 & 0.00 &-- & $-$0.01 & \\
BL-3 & 1859c2 & 18 35 27.640& \phantom{-}$-$34 31 59.353 & 15.53 & 4500 & 2.3 & $-$0.03 & 1.4 &  0.0 & $-$0.05 & 0.00 &-- & $-$0.02 &  \\
BL-4 & 3328c6 & 18 35 21.240& \phantom{-}$-$34 44 48.217 & 14.98 & 4700 & 2.0 & 0.13 & 1.5 & 0.00 & 0.00 & 0.00 & -- & 0.00 & \\
BL-5 & 1932c2 & 18 36 01.148& \phantom{-}$-$34 31 47.913 & 15.39 & 4500 & 2.1 & 0.16 & 1.6 &  0.00 & 0.00 & 0.00 &-- & 0.00 & \\
BL-7 & 6336c7 & 18 35 57.392& \phantom{-}$-$34 38 04.621 & 15.33 & 4700 & 2.4 & $-$0.47 & 1.4 &  $-$0.30  & $-$0.30 & $-$0.30 &$-$0.30 & $-$0.30 & \\

B3-b1 & 132160C4 & 18 08 15.971 & \phantom{-}$-$25 42 09.801&16.35& 4300 & 1.7 & $-$0.78 & 1.5 &  $-$0.30  & $-$0.30 & $-$0.90 &$-$0.30 & $-$0.35 & \\
B3-b2 & 262018C7 & 18 09 14.192 & \phantom{-}$-$25 56 47.300&16.63& 4500 & 2.0 & 0.18 & 1.5 &  --- & 0.00 & 0.00 &-- & 0.00 &   \\
B3-b3 & 90065C3 & 18 08 46.527 & \phantom{-}$-$25 42 44.401&16.59& 4400 & 2.0 & 0.18 & 1.5 & --- & 0.00 & 0.00 &-- & 0.00 & \\
B3-b4 & 215681C6 & 18 08 44.597 & \phantom{-}$-$25 57 56.802&16.36& 4500 & 2.1 & 0.17 & 1.7 & 0.00 & 0.00 & 0.00 &-- & 0.00 &  \\
B3-b5 & 286252C7 & 18 09 00.644 & \phantom{-}$-$25 48 06.699&16.23& 4600 & 2.0 & 0.11 & 1.5 &  0.00 & 0.00 & 0.00 & -- & 0.00 & \\
B3-b7 & 282804C7 & 18 09 16.670 & \phantom{-}$-$25 49 26.006&16.36& 4400 & 1.9 & 0.20 & 1.3 & 0.00 & 0.00 & 0.00 &-- & 0.00 &  \\
B3-b8 & 240083C6 & 18 08 24.733 & \phantom{-}$-$25 48 44.300&16.49& 4400 & 1.8 & $-$0.62 & 1.4 &  $-$0.10   & $-$0.30  & 0.00 &0.00 & $-$0.10 & \\

B3-f1 & 129499C4 & 18 08 16.301 & \phantom{-}$-$25 43 19.104&16.32& 4500 & 1.9 & 0.04 & 1.6 & 0.00 & 0.00 & 0.00 &--  & 0.00 &  \\
B3-f2 & 259922C7 & 18 09 15.730 & \phantom{-}$-$25 57 32.701&16.54& 4600 & 1.9 & $-$0.25 & 1.8 & 0.00   & 0.00  & 0.00 &-- & 0.00 & \\
B3-f3 & 95424C3 & 18 08 49.747 & \phantom{-}$-$25 40 36.898&16.32& 4400 & 1.9 & 0.06 & 1.7 &  0.00 & 0.00 & 0.00 & -- & 0.00 &  \\
B3-f4 & 208959C6 & 18 08 44.419 & \phantom{-}$-$26 00 25.001&16.51& 4400 & 2.1 & 0.09 & 1.5 &  0.00 & --- & 0.00 &-- & 0.00 & \\
B3-f5 & 49289C2 & 18 09 18.531 & \phantom{-}$-$25 43 37.403&16.61& 4200 & 2.0 & 0.16 & 1.8 &  0.00 & 0.00 & 0.00 &-- & 0.00 &  \\
B3-f7 & 279577C7 & 18 09 23.818 & \phantom{-}$-$25 50 38.104&16.28&  4800 & 2.1 & 0.16 & 1.7 &  0.00  & 0.00 & 0.00  &  &0.00 & \\
B3-f8 & 193190C5 & 18 08 12.757 & \phantom{-}$-$25 50 04.404&16.26& 4800 & 1.9 & 0.20 & 1.5 &  0.30 &0.00  &0.30  &--  & 0.20 & \\
BWc-1 & 393125 & $\phantom{-}$18 03 50.445 & $-$30 05 31.993&16.84& 4476 & 2.1 & 0.09 & 1.5  & 0.00 &0.00  &0.15  &-- & 0.05 & \\
BWc-2 & 545749 &$\phantom{-}$18 03 56.824 & $-$30 05 37.390&17.19& 4558 & 2.2 & 0.18 & 1.2   &0.10 &$-$0.30 &-$-$0.30 & -- & $-$ 0.16 &   \\  
BWc-3 & 564840 &$\phantom{-}$18 03 54.730 & $-$30 01 06.096&16.91& 4513 &2.1 &0.28 &1.3  & 0.00 & 0.00&0.00  & -- &0.06  & \\
BWc-4 & 564857 & $\phantom{-}$18 03 55.416 & $-$30 00 57.314&16.76& 4866 & 2.2 & 0.05 & 1.3   &0.10 &0.00  &0.00  &-- & 0.03  & \\
BWc-5 & 575542 &$\phantom{-}$18 03 56.021& $-$29 55 43.716&16.98& 4535 &2.1 &0.42 & 1.5   & 0.30 & 0.00  & 0.30  &-- & 0.20  &  \\
BWc-6 & 575585 &$\phantom{-}$18 03 56.543& $-$29 55 11.787&16.74& 4769 &2.2 &$-$0.25 &1.3 &$-$0.10 & 0.00 &$-$0.05 &$-$0.02 &-0.04 &  \\  
BWc-7 & 67577 &$\phantom{-}$18 03 56.543& $-$29 55 11.787& 17.01 & 4590 &2.2 &$-$0.25 &1.1 &$-$0.30 & -- & $-$0.30 &-- &$-$0.30&   \\   
BWc-8 & 78255 & $\phantom{-}$18 03 12.494& $-$30 03 59.111& 16.97 & 4610 &2.2 &0.37 &1.3   & 0.10 &$-$0.30  &0.00  &-- & $-$0.06  & \\  
BWc-9 & 78271 &$\phantom{-}$18 03 16.683& $-$30 03 51.406& 16.90 & 4539 &2.1 &0.15  &1.5  &0.20 &$-$0.05  &0.00  &-- & 0.05    & \\   
BWc-10 & 89589 &$\phantom{-}$18 03 18.914& $-$30 01 09.983& 16.70  & 4793 &2.2 & 0.07 & 1.3  &$-$0.10 & $-$0.10 &0.00  & -- & $-$0.06  & \\   
BWc-11& 89735 &$\phantom{-}$18 03 04.749& $-$29 59 35.301&  16.69  & 4576 &2.1 &0.17  & 1.0 & $-$0.10 & $-$0.10 & $-$0.10  &-- & $-$0.10  &  \\   
BWc-12& 89832 & $\phantom{-}$18 03 20.102& $-$29 58 25.785& 16.92  & 4547 &2.1 &0.23  &1.3 & 0.30 & 0.00 & 0.00  & -- & 0.10  & \\    
BWc-13& 89848 & $\phantom{-}$18 03 04.612& $-$29 58 14.080& 16.73  & 4584 &2.1 &0.36 &1.1   &0.30 & $-$0.30 &$-$0.15  &-- & $-$0.05 &  \\  

\noalign{\vskip 0.1cm}
\noalign{\hrule\vskip 0.1cm}
\noalign{\vskip 0.1cm}  
\hline                  
\end{tabular}
\end{flushleft}
\end{table*}  

\begin{table*}
\begin{flushleft}
\caption{Uncertainties on derived [Mn/Fe] for the metal-poor star B3-b1 and the red clump
star BWc-6. References: (1) Lecureur et al. (2007) (present results);
(2) Zoccali et al. (2008); (3) Ryde et al. (2010); (4) Hill et al. (2011) GIRAFFE data;
(5) Hill et al. (2011) UVES data (present results). 
[Mn/Fe] is also derived with changes in model parameters of $\Delta$T$_{\rm eff}$ = -150 K,
$\Delta$log g = 0.2, $\Delta$v$_{\rm t}$ = 0.1 km s$^{-1}$,  $\Delta$[Fe/H] = -0.1 dex.
 We note that star B3-b1
was chosen for its reanalysis by Ryde et al. (2010), but it has one discardable line that
gives a low Mn (in parenthesis) to which a low weight is given
 since it disagrees with the other three lines, and only
an extra -0.05 dex is added to the mean of the other three lines.
}             
\label{uncertainties}      
\centering          
\begin{tabular}{lrrrrrrrrrrr}     
\noalign{\smallskip}
\hline\hline    
\noalign{\smallskip}
\noalign{\vskip 0.1cm} 
Star & \hbox{T$_{\rm eff}$} & \hbox{log~$g$} & [Fe/H] & \hbox{v$_{\rm t}$} & ref. & [Mn1/Fe] & 
[Mn2/Fe] & [Mn3/Fe] & [Mn4/Fe] & [Mn/Fe] &  \\ 
\noalign{\vskip 0.1cm}
\noalign{\hrule\vskip 0.1cm}
\noalign{\vskip 0.1cm}  
B3-b1 &4300 & 1.7& -0.78& 1.5 & 1 &   -0.3  & -0.3 & (-0.9) &-0.3 & -0.35 &  \\
B3-b1 &4400 & 1.7& -0.60& 1.3 & 2 & -0.3 & -0.3 & (-1.0) & -0.40 & -0.38 &   \\
B3-b1 &4365 & 2.0& -0.73& 1.5 & 3 & -0.3 & -0.3 & (-0.9) & -0.35 & -0.37 &   \\
B3-b1 &4150 & 1.7& -0.78& 1.5 & -- & -0.4 & -0.45 & (-1.0) & -0.5 & -0.50 &   \\ 
B3-b1 &4300 & 2.2& -0.78& 1.5 & -- & -0.3 & -0.3 & (-0.8) & -0.3 & -0.35 &   \\ 
B3-b1 &4300 & 1.7& -0.78& 1.6 & -- & -0.3 & -0.3 & (-0.9) & -0.6 & -0.35 &   \\ 
B3-b1 &4300 & 1.7& -0.88& 1.5 & -- & -0.15 & -0.15 & (-0.7) & -0.15 & -0.20 &   \\ 
\noalign{\vskip 0.1cm}
\noalign{\hrule\vskip 0.1cm}
BWc-6 &4787 & 2.2 & -0.35 & 1.5 & 1 & 0.0 & +0.1 & 0.0 & 0.0 & 0.03 &   \\
BWc-6 &4769 & 2.2 & -0.17 & 1.2 & 4 & -0.05 & -0.05 & -0.1 & -0.1 &-0.08 &  \\
BWc-6 &4769 &2.2 &-0.25 &1.3 &5 & -0.1 & 0.0 &-0.05 &-0.02 &-0.04 &   \\
BWc-6 &4620 &2.2 &-0.25 &1.3 &-- & 0.0 & -0.3 &-0.2 &-0.25 &-0.19 &   \\ 
BWc-6 &4769 &2.7 &-0.25 &1.3 &-- & 0.0 & 0.0 &0.0 &0.0 &0.00 &   \\ 
BWc-6 &4769 &2.2 &-0.25 &1.4 &-- & 0.0 & 0.0 &0.0 &0.0 &0.00 &   \\ 
BWc-6 &4769 &2.2 &-0.35 &1.3 &-- &0.0  &0.0  &0.0 &0.0 &0.00 &   \\ 
 \hline
\noalign{\vskip 0.1cm}
\noalign{\hrule\vskip 0.1cm}                  
\end{tabular}
\end{flushleft}
\end{table*}  

\begin{table}[ht!]
\caption{Uncertainties on the derived [Mn/Fe] value for model changes of $\Delta$T$_{\rm eff}$ = -150 K,
$\Delta$log g = 0.2, $\Delta$v$_{\rm t}$ = 0.1 km s$^{-1}$,  $\Delta$[Fe/H] = -0.1 dex,
  and corresponding total error.
} 
\label{errors2}
\[
\begin{array}{lrrrrrr}
\hline\hline
\noalign{\smallskip}
\hbox{Star} &\hbox{[Fe/H]}& \hbox{$\Delta$T} & \hbox{$\Delta$$\log$ g}
& \hbox{$\Delta$[Fe/H]} & \hbox{$\Delta$v$_{t}$} & \hbox{($\sum$x$^{2}$)$^{1/2}$} \\
 & \hbox{($-$150 K)} & \hbox{(+ 0.2 dex)} & \hbox{($-$0.1 dex)} & \hbox{(+ 0.1 kms$^{-1}$}) &  \\
\hbox{(1)} & \hbox{(2)} & \hbox{(3)} & \hbox{(4)} & \hbox{(5)} & \hbox{(6)} \\

\noalign{\smallskip}
\hline
\noalign{\smallskip}
\noalign{\vskip 0.1cm}
\noalign{\hrule\vskip 0.1cm}
\hbox{B3-b1}  & $-$0.78        &  $-$0.09  &   +0.01 & +0.10 & $-$0.03 & 0.14 \\
\hbox{BWc-6} & $-$0.25       &   $-$0.17  &  +0.01 & +0.02  &  +0.02 & 0.17 \\
\noalign{\vskip 0.1cm}
\hline
\noalign{\smallskip}
\end{array}
\]
\end{table}

\begin{table}
\begin{flushleft}
\caption{Mean and dispersion of [Mn/Fe] in bulge stars in four bins of metallicity. 
The last column gives the number of stars in each bin.}         
\label{rms}      
\centering          
\begin{tabular}{lrrrrrrrrrrrrrr}     
\noalign{\smallskip}
\hline\hline    
\noalign{\smallskip}
\noalign{\vskip 0.1cm} 
\hbox{Bin} &  \hbox{$<$[Mn/Fe]$>$} & \hbox{$\sigma_{Mn/Fe}$}
 & \hbox{N$_{\rm stars}$}\\
\noalign{\vskip 0.1cm}
\noalign{\hrule\vskip 0.1cm}
\noalign{\vskip 0.1cm}
$\rm -0.8\leq [Fe/H]<-0.5$    &  -0.13    &    0.15  &   4\\
$\rm -0.5\leq [Fe/H]<-0.2$    &  -0.09    &    0.12  &  11\\
$\rm -0.2\leq [Fe/H]<+0.1$    &   0.00    &    0.03  &  16\\
$\rm +0.1\leq [Fe/H]<+0.4$    &   0.00    &    0.07  &  22\\
\noalign{\vskip 0.1cm}
\noalign{\hrule\vskip 0.1cm}
\noalign{\vskip 0.1cm}  
\hline                  
\end{tabular}
\end{flushleft}
\end{table}

\section{Results}

Table \ref{atmos1} reports the Mn abundances derived in the present work.
In Fig. \ref{plotbojo}a
our results are plotted with different symbols identifying
each of the four fields and the red clump stars.
 The first conclusion that can be drawn is that there is no clear 
distinction in Mn abundances among the four bulge fields.
Figure \ref{plotbojo}  clearly indicates 
that,  from values of [Mn/Fe] $\sim$-0.7 at [Fe/H]$\sim$-1.3, 
[Mn/Fe] increases to the solar value at the solar metallicity,
 although in the metallicity range
 -0.7$\simless$[Fe/H]$\simless$-0.2 the trend is somewhat ill-defined because of the 
scarcity of data points and their relatively large dispersion.
To understand whether this dispersion is true, in Fig. \ref{movingrms} we show 
our results for the bulge stars together with a moving average line 
(moving average on 10 points) and the corresponding rms dispersion in each subsample 
(shown as dashed lines). This plot indicates a larger dispersion of [Mn/Fe] for
 metallicities below solar, and even larger for [Fe/H]$<-0.5$,
 although in this last part the scarcity of data may be the reason for the apparent increased dispersion (a large [Fe/H] range enters the first few moving average samples, so that the slope in Mn/Fe acts to increase the dispersion among the sample). 
Table \ref{rms} shows
the mean [Mn/Fe] values in four metallicity bins, 
together with their dispersion, and confirms that [Mn/Fe]
 may be more dispersed at subsolar metallicities.
%


\subsection{Comparison with chemical evolution models}

Chemical evolution model predictions by Timmes et al. (1995) and
 Tsujimoto \& Shigeyama (1998, hereafter TS98)  for the solar neighborhood,
as well as Cescutti et al. (2008) for the Galactic bulge
are also displayed in Fig. \ref{plotbojo}. 

Timmes et al. (1995) carried out computations of chemical evolution models 
 for the solar neighborhood, based on
 metallicity dependent yields by WW95. The nucleosynthesis results 
by WW95 consisted of 60 type II supernova models of masses 
11 $\simless$ M/M$_{\odot}$ $\simless$ 40
and metallicities from zero to solar, incorporating 76 stable isotopes from H to Zn.
Timmes et al. also included yields from type Ia SNe, 
Big-Bang nucleosynthesis of
the light elements, and mass loss from planetary nebulae.
For SNe Ia, Timmes et al. followed the prescriptions by 
Matteucci \& Greggio (1986), adopting the results from SNe Ia models by
Nomoto et al. (1984) and the yields from Thielemann et al. (1986).
No explicit dependence on metallicity was employed for the SNe Ia input data.
 
Tsujimoto \& Shigeyama (1998) employed the SNe II yields from Nomoto et al. (1997) and the chemical 
evolution model by Yoshii et al. (1998).
 There is no particular consideration of yields as a function
of metallicity.
The TS98 models imply low [Mn/Fe] values for metallicities
[Fe/H]$\simless$-2.8, and a Mn plateau at -2.8$\simless$[Fe/H]$\simless$-1.0,
because the most massive stars eject about 2 orders of magnitude
larger mass of Mn than do the lower mass SNe II.
 For [Fe/H]$>$-1 the 
SN Ia cause a clear increase in the [Mn/Fe] values.
Both Timmes et al. and TS98 models assume a Salpeter
initial mass function (IMF).

 Cescutti et al. (2008) computed models specifically
for the Galactic bulge,
by assuming a flatter IMF and a star formation efficiency 20 times higher
 than the solar vicinity, as previously established by Ballero et al.
(2007). Their most suitable model option involves
 metallicity dependent yields from WW95 for
 type II SNe, combined with metallicity dependent yields from
type Ia SNe, that were obtained by
 applying a dependence law
proportional to the overall metallicity Z, of Z$^{0.65}$,
 to yields from Iwamoto et al. (1999).

\subsection{Comparison with bulge Mn abundances from the literature}

In Fig. \ref{plotbojo}b the present results are compared with
 literature Mn abundances
for bulge giants, including bulge field stars  by 
McWilliam et al. (2003a,b)\footnote{Mn abundances reported by McWilliam et al. (2003a,b) 
for the Sagittarius dwarf galaxy are not considered here, 
since this galaxy has a very different star formation history.}, and three stars
of NGC 6528, with stellar parameters derived by Zoccali
et al. (2004), and Mn abundances derived by Sobeck et al. (2006)
from the same spectra.
The present Mn abundances appear to be in very good agreement
with these data points both for field bulge giants 
(McWilliam et al. 2003a,b),
and for the bulge globular cluster stars (Sobeck et al. 2006).

These data, together with the present results confirm a good agreement
with the chemical evolution model by Cescutti et al. (2008). On the other
hand the  lack of a well-defined [Mn/Fe] trend in the range
-0.7$\simless$[Fe/H]$\simless$-0.2, or even up to [Fe/H]$\sim$0.0 if
the McWilliam et al. field stars and NGC 6528 data (Sobeck et al.) are
taken into account,
does not allow the exclusion of the behaviour indicated in the models by TS98 and
Timmes et al. (1995).

{\it Comparison with the galactic thin disc, thick disc, and halo}

 In Figs. \ref{lit} and \ref{plotsobeck} our results are plotted again,
 with different symbols identifying
the four fields and the Baade's Window red clump stars (as in Figs. \ref{plotbojo}a,b), 
now compared with literature data, for the thin disk, thick disk,
and halo stars.


In Fig. \ref{lit}a, a
 comparison with Mn abundances in thin and thick disc stars by
 Feltzing et al. (2007) indicates that the bulge giants show good agreement
with both thin and thick disc stars for metallicities higher than 
[Fe/H]$\simgreat$-0.7, whereas for lower abundances the [Mn/Fe] values in the bulge
appear to decrease faster than for the few thick disc stars with 
[Fe/H]$<$-0.7.
The comparison with thin and thick disc stars by Reddy et al. (2003, 2006)
given in Fig. \ref{lit}b shows a decrease in [Mn/Fe] for the thick disc 
stars that is more consistent with that
observed for our bulge sample despite a rather large spread.

Figure \ref{lit}c  shows a mean [Mn/Fe]$\sim$-0.4 abundance in
 very  metal-poor
halo stars, from Cayrel et al. (2004), and [Mn/Fe]$\sim$-0.3,
 in moderately metal-poor halo stars from Nissen et al. (2011).
 A slope between the most metal-poor and the
moderately metal-poor halo stars could be present.
 The two stars BW-f4 and BW-f8, at [Fe/H]$\sim$-1.3, have [Mn/Fe]$\sim$-0.7.
However, they have  S/N = 22 and 38, respectively, in the original red spectra
(Lecureur et al. 2007), therefore not as high as needed to claim
a difference compared to other samples. In the star B3-b1 at [Fe/H]$\sim$-0.8,
 one line indicates
a low Mn abundance, in contrast to the other three lines which show 
moderate Mn deficiencies; therefore, we gave a low weight to the discrepant line,
only lowering the mean by -0.05 dex.

In Fig. \ref{plotsobeck} our results are compared with those by 
Ishigaki et al. (2013), Sobeck et al. (2006), Gratton (1989), and
Prochaska et al.
(2000), adopting for the last their mean values for standard model atmospheres.
Ishigaki et al. (2013) derived the abundances of iron-peak elements including Mn
in 97 thick disk, inner and outer halo stars. Their [Mn/Fe] ratios
 (Fig. \ref{plotsobeck}a)
 start to drop from the solar value at [Fe/H]$\sim$-0.7, down to
 [Mn/Fe]$\sim$-0.3 in the range -1.6$\simless$[Fe/H]$\simless$-1.1, 
and drop lower for the more metal-poor stars. The
data could be interpreted as a steady increase
 from [Mn/Fe]$\sim$-0.6 at [Fe/H]$\sim$-3.0 up to
[Mn/Fe]$\sim$0.0 at [Fe/H]$\sim$-0.7.
 The pace of decrease of [Mn/Fe] runs in parallel
with our bulge metal-poor stars, but with higher values.
Sobeck et al. (2006) derived Mn abundances for a large number of field stars,
and stars in globular clusters, and in one old open cluster. 
In particular for 200 stars in 19
globular clusters in the metallicity range -2.7 $<$ [Fe/H] $<$ -0.7, 
an approximately constant value is found with a mean [Mn/Fe]=-0.37.
     Somewhat lower values of [Mn/Fe]$\sim$-0.4 were found for field stars
by Gratton (1989) (Fig. \ref{plotsobeck}c), Cayrel et al. (2004) 
(Fig. \ref{lit}c), and Ishigaki et al. (2013) (Fig. \ref{plotsobeck}a),
and [Mn/Fe]$\approx$-0.5 by Reddy et al. (2006) (Fig. \ref{lit}b). 
A few stars at metallicities -1.6$<$[Fe/H]$<$-1.0 show [Mn/Fe]$\sim$-0.6,
compatible with the results for our bulge stars.
Prochaska et al. (2000) analysed ten thick disc stars (Fig. \ref{plotsobeck}c)
in the metallicity range -1.2 $\simless$ [Fe/H] $\simless$ -0.4. 
Their [Mn/Fe]$\sim$-0.3
at [Fe/H]$\sim$-0.4 drops to [Mn/Fe]$\sim$-0.4 at [Fe/H]$\sim$-1.2, in good
agreement with the present results for the metal-poor bulge giants. 

Del Peloso et al. (2005) derived Mn abundances for 20 thin disc stars, with
results compatible with those by Feltzing et al. (2007) and Reddy et al.
(2003). Allen \& Porto de Mello (2011) derived Mn abundances in barium stars.

 A further inspection of the data can be obtained by considering oxygen
instead of iron as reference element, given that oxygen is a bona fide
primary element produced uniquely by SNe II.
The same applies to magnesium, however there is a possibility that
some magnesium is produced in SNe Ia. As a matter of fact, Feltzing
et al. (2007) were able to better disentangle thick disc from thin disc stars,
through this approach. The oxygen 
 abundances for the present sample of stars
were derived in Zoccali et al. (2006) and Lecureur et al. (2007),
with uncertainties of $\pm$0.15 in [O/H].
 These data are used to plot [Mn/O] vs. [Fe/H] and
vs. [O/H] (Figs. \ref{mno1}a,b). 
These figures disentangle the stellar populations more effectively, showing that the present
bulge stars have a [Mn/O] behaviour similar to that of the thick disk, extending the
 thick disc trend to higher metallicites (both in [Fe/H] and [O/H]), 
and distinct from the thin disc trend at all metallicities. 
In terms of Mn production, Fig. \ref{mno1}b has interesting implications since 
it should show more directly the metallicity dependence of Mn production, as well 
as the SNe Ia production of Mn.
Cescutti et al. (2008) claim that both SNe II and SNe Ia 
metallicity dependent yields are needed to explain the Mn chemical evolution of the 
solar neighborhood, the Galactic bulge, and the Sgr dSph; if this were the case, one 
would typically expect two regimes of chemical enrichments: (i) dominated by 
SN II 
at low metallicities, possibly with a slope of [Mn/O] with [O/H] reflecting the metallicity 
dependence of the SNe II yields, followed by (ii) an enrichment dominated by SNe Ia, possibly 
with a slope reflecting the metallicity dependence of the SNe Ia yields.
North et al. (2012) discuss these contributions further in the context of
dwarf spheroidal galaxies (see their Fig. 7). The transition 
between the two regimes should occur at different [O/H] in the different populations, 
mirroring the decrease in [$\alpha$/Fe] at the metallicity reached by
 the system when the 
delayed SNe Ia yields enter into play in the chemical evolution. The slope in regime (ii)
 can also depend on the 
population since the SNe Ia contributing at a given [O/H] can be of different metallicities. 
Given that $\alpha$-elements evolution of $\alpha$-elements in thick disc and
bulge stars is similar (e.g. Alves-Brito et al. 2010; 
Gonz\'alez et al. 2011, Hill et al. 2011), it is hence expected that the bulge follows the 
thick disc in regime (ii) when SNe Ia are contributing, i.e. up to metallicities 
of [O/H]$\sim 0.0$. 
In this same metallicity range, the thin disc already includes SNe Ia products and 
[Mn/O] is hence larger. 
Beyond solar [O/H], if the thick disc and bulge differ, it could
 be partly attributed to
 the metallicity dependent yields of SNe Ia. 
 We note that the mean [Mn/O] in the bulge does not continue to
 rise for [O/H]$>$0.0, as one could expect if the SNe Ia yields were strongly 
metallicity dependent.
According to Cescutti et al. (private communication), the sensitivity of bulge models 
to the SNe Ia metallicity dependency is, however, negligible, at least when
 the SNe Ia progenitors are formed with roughly solar metallicities,
and using yields from Maeder (1992) for oxygen, in which it is possible that stellar 
rotation decreases the oxygen yields possibly too extremely.
 It would be very interesting to compare the Mn/O in the bulge and 
thick disc to chemical evolution
 models with various hypotheses on the metallicity dependency 
of SNe Ia yields, and different models of oxygen yields, 
such as those of Cescutti et al. (2008). 

In conclusion,
in the metal-poor range -1.3 $<$ [Fe/H] $<$ -0.8, 
 the present sample includes two stars that are
 slightly more Mn-poor than shown in the literature, except for a closer agreement with
data from Prochaska et al. (2000) and Reddy et al. (2006). 
 The increase towards solar abundances starts to occur at around
-1.0$\simless$[Fe/H]$\simless$-0.7, reaching the solar value
at solar metallicity, but showing a spread of -0.3$<$[Mn/Fe]$<$0.0
in the range -0.7$<$[Fe/H]$<$-0.2. We note that a spread is also seen
in some dwarf spheroidals (North et al. 2012), and could be explained
if two channels of SNe Ia exist (G. Cescutti, private communication).
The behaviour of [Mn/O] vs. [O/H] (Fig. \ref{mno1}b) seems to indicate that the present
sample of bulge stars have had a different chemical enrichment history
to thick disc stars.

\section{Discussion}

A chemical evolution model by Tsujimoto \& Shigeyama (1998)
predicted that [Mn/Fe] would decrease with decreasing metallicity for
[Fe/H]$\simless$-3.0, and would otherwise be a constant [Mn/Fe]$\sim$-0.4
in the range     -3.0$\simless$[Fe/H]$\simless$-1.0, and would then increase
to the solar value reached at solar metallicity.
The present and literature results agree with these predictions of low
[Mn/Fe] in the halo, possibly with a slope, that 
starts to increase due to the contribution of SNe Ia at 
-1.0$<$[Fe/H]$\sim$-0.7.

Timmes et al. (1995) presented calculations of Mn production
as a function of metallicity
through production in massive stars, with a metallicity dependent yield
due to the lower neutron excess in the metal-poor stars,
 whereas SNe Ia contribute significantly to
the Mn abundance at higher metallicities. Their models predicted a lower
constant value in the halo of [Mn/Fe]$\sim$-0.6, and the increase in
[Mn/Fe] from [Mn/Fe]$\sim$-0.6 at [Fe/H]$\sim$-1.0 to the solar ratio at
solar metallicity, in good
 agreement with the present results for the bulge stars. 

 The Galactic bulge models by Cescutti et al. (2008), that include
metallicity dependent yields for both SNe II (WW95) and SNe Ia, appear
to most accurately reproduce the present results, confirming previous work
by McWilliam et al. (2003).

The present results confirm conclusions from previous work, namely,
 a low [Mn/Fe] in the halo and metal-poor thick disc stars
with [Fe/H]$\simless$-0.7 indicating that
these stars were enriched in Mn by type II SNe. The contribution from
SN Ia starts operating at  -1.0$\simless$[Fe/H]$\simless$-0.7, 
possibly being different for the bulge, thick disk, and thin disk,
given some differences shown in Figs. \ref{lit}, \ref{plotsobeck},
and \ref{mno1}.

\section{Conclusions}

Abundances of iron-peak elements in different environments
 are key ingredients constraining nucleosynthesis
yields in type II SNe and type Ia SNe, as well as in chemical evolution models
(e.g. Iwamoto et al. 1999). The relative abundances of several iron-peak elements
(species) relative to iron in the Galactic halo, disk, and bulge, as well as in
globular clusters and dwarf galaxies can constrain models.
The behaviour of [Mn/Fe] vs. [Fe/H] shows that the iron-peak element Mn
has not been produced in the same conditions as even-Z iron-peak elements
such as Fe and Ni. 

Abundances of Mn in metal-poor halo stars show 
 [Mn/Fe]$\sim$-0.4 in the metallicity range -3.0$\simless$[Fe/H]$\simless$-1.5
(Cayrel et al. 2004; Ishigaki et al. 2013)
confirming its production in core collapse SNe,
similarly to the $\alpha$ and heavy elements.
As noted by Gratton (1989) [Mn/Fe]$\approx$-0.4 show a symmetric
behaviour relative to the [$\alpha$/Fe]$\approx$+0.4 in the halo.
 The metallicity-dependent yields
from WW95 essentially reproduce this behaviour.

The triggering of type Ia SNe seems to occur at 
around -1.0$\simless$[Fe/H]$\simless$-0.7 and predicted
yields reproduce satisfactorily the steady increase of [Mn/Fe] 
in thick disc stars,
reaching the solar value at around 
-0.7$\simless$[Fe/H]$\simless$-0.0, as shown in Figs. \ref{lit} and
\ref{plotsobeck}. We note that the data by Feltzing et al. (2007) show higher abundances
of [Mn/Fe]$\sim$-0.2 in metal-poor thick disc stars, whereas data by Gratton (1989)
and Prochaska et al. (2000) reach the solar [Mn/Fe] ratio at
metallicities of [Fe/H]$\sim$-0.2.

Cescutti et al. (2008) were able to reproduce the behaviour of [Mn/Fe] vs. [Fe/H]
for the seven bulge stars by McWilliam et al. (2003) by adopting a metallicity
dependent yield for massive stars (WW95) and
 introducing a metallicity dependency in the yields by Iwamoto et al. (1999)
proportional to Z$^{0.65}$. We find that their model is the most suitable to fit the present data.
On the other hand, given a spread in Mn-over-Fe abundances in the metallicity range 
 -0.7$\simless$[Fe/H]$\simless$-0.2, the higher Mn abundances at [Fe/H]$\sim$-0.7 indicated
in the TS98 model, or lower values as predicted by Timmes et al. (1995) cannot be excluded.

 Using oxygen as reference element, the [Mn/O] trend runs parallel 
to the thick
disc stars, at the bulge higher metallicities. However, at the highest [O/H], 
the [Mn/O] in the bulge stars does not seem to increase anymore, possibly indicating that
 the SNe Ia metallicity dependency is weak, or even absent. 
We caution, however, that the measurement of [Mn/O] suffers from large errors in these stars
 (mostly because of the O abundance uncertainties).

Finally, the low Mn abundance at [Fe/H]$\sim$-1.0 is in good agreement 
with results from damped Lyman-$\alpha$ systems 
(e.g. Pettini 1999; Fig. 24 in Prochaska et al. (2000)).


\begin{acknowledgements}
      We are grateful to G. Cescutti for useful comments on
chemical enrichment.
      BB acknowledges partial financial support by CNPq and FAPESP.
MZ and DM are supported by Fondecyt Regular 1110393, and 1130196, the BASAL Center for
Astrophysics
and Associated Technologies PFB-06, the FONDAP Center for Astrophysics
15010003,
Proyecto Anillo ACT-86, and the Chilean Ministry for the Economy,
Development, and Tourism Programa Iniciativa Cient\'{\i}fica Milenio through grant
P07-021-F, awarded
to The Milky Way Millennium Nucleus. MT acknowledges FAPESP postdoctoral
fellowship no. 2012/05142-5.

\end{acknowledgements}

\end{document}